\documentclass[preprint]{revtex4}
\usepackage{graphicx}
\usepackage{times}
\usepackage{dcolumn}
\begin{document}

\title{\bf Modeling the selective partitioning of cations into negatively charged 
nanopores in water}

\author{Lu Yang$^{1,2,*}$ and Shekhar Garde$^1$\footnote[1]{e-mail: 
gardes@rpi.edu, yangl@lanl.gov}}

\affiliation{$^1$The Howard P. Isermann Department of Chemical \&
Biological Engineering, and Center for Biotechnology \&
Interdisciplinary Studies, Rensselaer Polytechnic Institute, Troy, NY
12180.\\ $^2$ T-12 Group, Los Alamos National Laboratory, Los Alamos,
NM 87545}

\date{\today}


\begin{abstract}
{Partitioning and transport of water and small solutes into and
through nanopores is important to a variety of chemical and biological
processes and applications.  Here we study water structure in
negatively charged model cylindrical [carbon nanotube (CNT)-like]
nanopores as well as the partitioning of positive ions of increasing
size ($Na^+$, $K^+$, and $Cs^+$) into the pore interior using
extensive molecular dynamics simulations.  Despite the simplicity of
the simulation system -- containing a short CNT-like nanopore in water
carrying a uniformly distributed charge of $q_{pore}=-ne$ surrounded
by $n$ ($= 0,\ldots , 8$) cations, making the overall system
charge-neutral -- the results provide new and useful insights on both
the pore hydration and ion partitioning.  For $n=0$, that is, for a
neutral nanopore, water molecules partition into the pore and form
single-file hydrogen-bonded wire spanning the pore length. With
increasing $n$, water molecules enter the pore from both ends with
preferred orientations, resulting in a mutual repulsion between
oriented waters at the pore center, and creating a cavity-like low
density region at the center. For low negative charge densities on the
pore, the driving force for partitioning of positive ions into the
pore is weak, and no partitioning is observed.  Increasing the pore
charge gradually leads to partitioning of positive ions into the pore.
Interestingly, over a range of intermediate negative charge densities,
nanopores display both thermodynamic as well as kinetic selectivity
toward partitioning of the larger $K^+$ and $Cs^+$ ions into their
interior over the smaller $Na^+$ ions. Specifically, the driving force
is in the order $K^+>Cs^+>Na^+$, and $K^+$ and $Cs^+$ ions enter the
pore much more rapidly than $Na^+$ ions.  At higher charge densities,
the driving force for partitioning increases for all cations -- it is
highest for $K^+$ ions, and becomes similar for $Na^+$ and $Cs^+$
ions. The variation of thermodynamic driving force and the average
partitioning time with the pore charge density together suggest the
presence of free energy barriers in the partitioning process. We
discuss the role of ion hydration in the bulk and in the pore interior
as well as of the pore hydration in determining the barrier heights
for ion partitioning and the observed thermodynamic and kinetic
selectivity.}
\end{abstract}
\maketitle 

\section{Introduction}
Understanding thermodynamics and kinetics of partitioning of small
molecules into nanoscopic pores or confined spaces of molecular
dimensions is important to a wide range of chemical and biological
processes and applications \cite{sholl:06:science, hummer:01:nature,
eijkel:05:microfluidics,holt:06:science}.  The pores of interest can
be smooth cylindrical (carbon nanotube-like) \cite{ajayan:93:nature}
or slit-pores (between flat surfaces) \cite{gelb:99:ropip}, or may
have complex geometries as observed in biological pores
\cite{agre:00:nature, doyle:science:98}, in activated carbon, and in
other nanoporous materials \cite{raman:96:cm}. When exposed to
solution, such materials can selectively extract and/or allow
transport of specific solute molecules into and through their pores,
depending on the pore-solvent, pore-solute, and solute-solvent
interactions.  Successful bottom-up design of such selective systems
can be aided by fundamental understanding of the interplay of these
interactions in idealized model systems.  Molecular dynamics
simulations are an excellent tool to build and simulate such model
systems aimed at fumdamental understanding.

Here we are interested in partitioning of simple spherically
symmetric ionic solutes into interior of charged hollow cylindrical
pores of well defined dimensions in the nanometer range. Such a model
system is relevant not only to electrostatically driven molecular
separations or nanofluidic applications, but also to development of
energy storage devices, such as carbon supercapacitors, that use ion
adsorption on the surface of highly porous materials to store charge.
A recent experimental study by Chmiola et al. \cite{chmiola:06:science}
shows that carbide-derived carbon materials with pore sizes from 0.6
to 2.25 nm when used as a negative electrode, can partition
tetraethylammonium cations ($\sim$0.68 nm in diameter) into the pore.
The closer approach of partially dehydrated cations to the nanoporous
electrode surface leads to significantly enhanced capacitance of these
materials.

Our simulation system contains a nanopore (modeled by a carbon
nanotube-like cylindrical pore) in water carrying a uniformly
distributed charge of $q_{pore}=-ne$ surrounded by $n$ ($= 0,\ldots ,
8$) cations (either $Na^+$, $K^+$, or $Cs^+$), making the overall
system charge-neutral.  We perform extensive nonequilibrium
simulations focused on the kinetics of partitioning, which show that
negatively charged nanopores can be selective toward partitioning of
the larger $K^+$ and $Cs^+$ ions and can exclude the smaller $Na^+$
ions over a range of nanopore charge densities.  We complement those
simulations with equilibrium free energy calculations of ion hydration
in bulk water and in the hydrated nanopore interior.  These
calculations along with analysis of ion hydration shell fluctuations
highlight the role of free energy barriers as well as equilibrium free
energy of partitioning which together are expected to govern the rate
of ion transport through selective pores.

\section{Simulation Details}
{\it Simulations of ion partitioning into nanopores}: We used a piece
of (5,5) armchair carbon nanotube (CNT)-like pore comprising 100
carbons [with the pore diameter ({\it i.e.}, C-C distance) of 6.7 \AA\
and length 11 \AA] as a model cylindrical pore.  A total charge of
$q_{pore}=-ne$ (where $e$ is the magnitude of electronic charge) was
distributed uniformly on the pore atoms, such that each atom carries a
charge of -$ne/100$.  The nanopore was placed in a solution containing
500 explicit water molecules and $n$ cations, thus, making the overall
system electrically neutral.  Note that our goal is not to model
specifically a CNT, but use its cylindrical geometry as a model for a
cylindrical pore.  A realistic model of CNT would need to consider the
chemistry of rim atoms, non-uniform distribution of charges (in case
of charged nanotubes) \cite{Keblinski:prl:02}, etc.

The present systems contain $n$ cations, but no explicit anions (the
pore with its $-ne$ charge acts as a large anion).  In realistic
systems, anions (e.g., chloride ions) will be present in the salt
solution. For $n=5$, we simulated the present system with excess salt
({\it i.e.}, with additional few pairs of NaCl, KCl, or CsCl) ions.
The results on partitioning of cations in this system were similar to
the ones reported here, suggesting that the trends on partitioning
reported here will be unaffected.

Simulations were performed using AMBER6.0 \cite{amber:95}, using TIP3P
\cite{TIP3P:83} model of water, and Lennard Jonesium (LJ) description
of carbon (atom type CA) \cite{amber:95, kalra:04:jpc}. Ions were
represented as LJ spheres with ion charge placed at the center
\cite{Straatsma:88}: $\sigma_{Na}=2.530$ \AA, $\epsilon_{Na}=0.06184$
kJ/mol, $\sigma_{K} =5.874 $ \AA, $\epsilon_{K} =0.0000568$ kJ/mol,
and $\sigma_{Cs}=6.0492$ \AA, $\epsilon_{Cs}=0.0003372$ kJ/mol.
Ion-water and ion-carbon LJ interactions were calculated using
Lorentz-Berthelot mixing rules \cite{allen87}.  Periodic boundary
conditions were applied and the particle mesh Ewald method
\cite{PME:93} was used to calculate the electrostatic interactions
with a grid spacing of 1 \AA. Temperature and pressure were maintained
at 300 K and 1 atm, respectively, using the Berendsen method
\cite{Berendsen:84}. A time step of 1 fs was used in all the 
simulations.

{\it Selective partitioning of cations into nanopores}: A given system
contains a carbon nanotube-like pore carrying a total charge of
$q_{pore}=-ne$, water, and $n$ cations.  We monitored the number of
cations partitioned into the nanopore as a function of time for
different values of $n$ (see Figure \ref{fig:selectivity}). Separate
simulations were performed for $q_{pore}=-1e$ through $-8e$ in steps
of $1e$, for each cation.  At the beginning of each production run
(marked $t=0$), each cation in the system was at least 9 \AA\ away
from the atoms of the model nanopore.  To obtain an estimate of the
partitioning time (or the first passage time for partitioning), each
simulation was run for a sufficiently long time.  For lower charge
densities on the nanopore, the electrostatic driving force is low, and
extended simulations of neutral and near neutral ($q_{pore}=-1e$)
nanopores show that ions are excluded from the pore interior,
consistent with previous studies \cite{kalra:03:pnas, hummer:05:bj,
sinnott:99:nt}.  For higher charge densities on the pore, not only is
the electrostatic driving force higher, but the ion concentration
outside the tube is slightly higher as well (by the way the system is
constructed), which likely further increases the driving force. Thus,
with increasing $n$ we expect ions to partition into the pore, with
the time required for partitioning decreasing with the pore charge
density.  Correspondingly, for $|q_{pore}| \ge 2e$ simulations were
performed for 24/$n$ nanoseconds each.

\section{Results and Discussion}

\begin{figure}[tbph]
\includegraphics[width=6.0in]{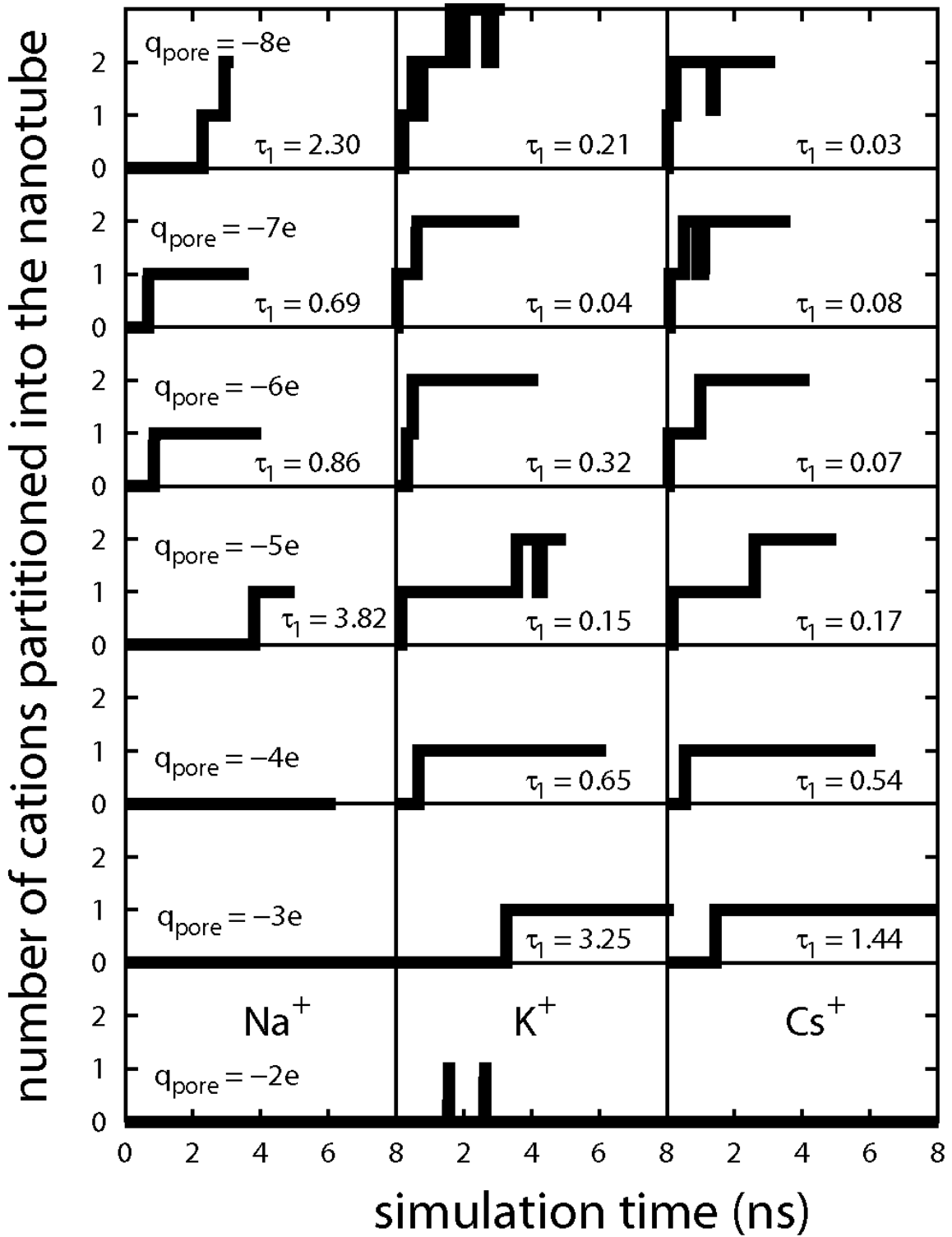}
\caption{Kinetics of ion partitioning from nonequilirbium MD simulations. 
Number of cations partitioned into the negatively charged nanopore in
water as a function of time in selected simulation runs. Nanopore
carries a charge of $-ne$, $n=2,\ldots 8$ from bottom to top. $\tau_i$
is the incremental time for the $i$-th cation to partition into the
pore. Data for $Na^+$ (left), $K^+$ (middle), and $Cs^+$ (right
column) are shown.}
\protect\label{fig:selectivity}
\end{figure}

{\it Kinetics of ion partitioning}: Figure \ref{fig:selectivity} shows
the number of cations partitioned into the pore for various charge
states of the pore observed in different nonequilibrium simulation
runs. For $q_{pore}=-2e$, the charge density on the nanopore is
sufficiently small and none of the ions partitions into the pore over
the timescale of 12 ns.  It is only for $q_{pore} = -5e$ we observe
partitioning of one $Na^+$ ion into the pore in the specific set of
runs shown in Figure \ref{fig:selectivity}.  In contrast, $Cs^+$ and
$K^+$ ions partition into the pore for $|q_{pore}| \geq 3e$.  At
higher values of $|q_{pore}|$, we observe partitioning of the second
and third (in case of $K^+$) cation into the charged nanopore.  To
obtain more quantitative estimates of the kinetics of partitioning, we
performed 10 independent nonequilibrium simulations for each ion type
starting with different initial configuration for $q_{pore}=-4e,\
-5e,\ {\rm and}\ -6e$. Each of these simulations were run until the
first cation partitions well into the pore, which for $q_{pore}=-4e$
was longer than 25-30 ns each for several runs with $Na^+$ ions.  We
define $\tau_1$ as the waiting time for the first cation to partition
into the nanopore.  Table \ref{tab:tau} lists average, $\left< \tau_1
\right>$, and standard deviation, $\sigma_{\tau_1},$ of the waiting
time distribution obtained from 10 simulations.  Although a larger
number of simulations may be needed to obtain more accurate estimates
of waiting time distributions, trends in our data are already clear
and interesting.  For example, for $q_{pore}=-4e$, $Na^+$ ion takes
$\sim$25 times longer to partition into the pore ($\sim$18 ns)
compared to that for $K^+$ and $Cs^+$ ions ($\sim$0.7-0.8 ns).  As the
charge on the nanopore is increased, partitioning of $Na^+$ occurs
over shorter timescales, and correspondingly, the $\tau_1$ values for
both $K^+$ and $Cs^+$ ions are small and decrease somewhat further.
Thus, from kinetic perspective, over a range of charge densities, the
nanopore is selective toward partitioning of the larger ions $K^+$ and
$Cs^+$ over the smaller ion $Na^+$, and increasing the driving force
significantly ({\it i.e.}, increasing $|q_{pore}|$) appears to
decrease that selectivity somewhat. To understand the ion partitioning
at a more fundamental level, we investigate below the structural and
thermodynamic aspects of the partitioning process.

\begin{figure}[tbph]
\includegraphics[width=6.0in]{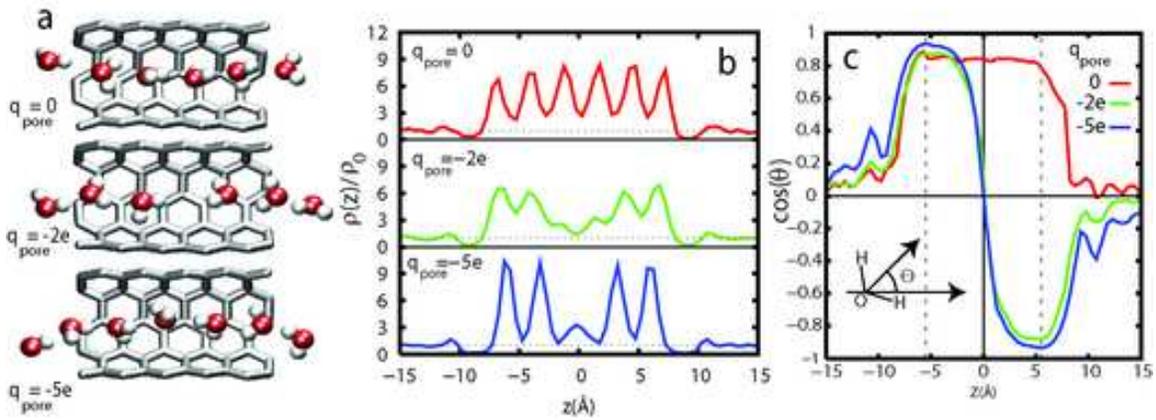}
\caption{Water structure in neutral and charged nanopores. (a) Snapshots
of water molecules, (b) local densities, and (c) orientations of water
molecules in the interior of neutral and charged nanopores carrying a
total charge of $-2e$ and $-5e$, respectively. $\theta$ is the angle
between dipole vector and the $Z$ axis as shown schematically in panel
(c).}
\protect\label{fig:density}
\end{figure}

{\it Water structure in neutral and charged pores}: Filling and
emptying transitions of water into neutral nanotubes have been studied
previously \cite{hummer:01:nature,waghe:02:jcp}. For the carbon-water
LJ parameters used here, water molecules partition into the tube and
form a single-file hydrogen bonded wire, in which each water molecule
donates and accepts one hydrogen bond to and from its neighbors on the
left and right (Figure \ref{fig:density}a).  The length of the tube is
such that only four water molecules completely fit inside the tube.
However, the hydrogen-bonding interactions of the boundary water
molecules with those inside the tube are sufficiently strong. As a
result, one water molecule on each side of the tube also maintains its
position and orientation consistent with those in the tube.
Correspondingly, the water density profile along the tube axis shows
six clearly defined peaks (Figure \ref{fig:density}b).

Water molecules in the hydrogen-bonded wire have one of their OH bond
vectors aligned approximately with the pore axis and pointing in the
positive $Z$ direction.  Properties of such hydrogen bonded wires have
been studied especially for their ability to transport protons at fast
rates \cite{hummer:03:prl}. The dipole vector OM of each water
molecule makes an angle of $\sim$35 degrees with the pore axis
[$\left< cos\theta \right>\approx 0.82$] (Figure \ref{fig:density}c).
We note that the mirror image of the chain with all OH vectors
pointing in the negative $Z$ direction is equally likely due to
two-fold symmetry \cite{waghe:02:jcp}.

The effects of charging the nanopore on the structure of water inside
the pore are interesting. For $q_{pore}=-2e$, that is, when the charge
on each carbon is $-0.02e$, water molecules enter the pore with both
of their hydrogen atoms pointing inward and the HOH planes roughly
parallel to the pore axis (Figure \ref{fig:density}a)
\cite{waghe:02:jcp}.  As the chains from the two ends meet at the
center, there is significant electrostatic repulsion between them.  As
a result, water density is depressed at the center of the pore as
indicated by the significant reduction in the heights of the two
central peaks (Figure \ref{fig:density}b). Water is effectively
expelled from that region, creating a molecular scale void or a
cavity.  This expulsion in similar to that observed recently in
simulations of charged plates in water \cite{rasaiah:05:jpc}. The
other peaks are also less well defined. As the charge on the nanopore
is increased, the stronger water-nanopore interactions lead to a
partial filling of the internal cavity. In the interior of these
charged nanopores, water molecules flip the orientation as we traverse
the pore from left to right as indicated by the $\left< cos\theta
\right>$ profile (Figure \ref{fig:density}c). A similar change in 
orientation of water molecules is observed in the aquaporin protein
channel, which has positively charged regions at the opening and the
exit to block transport of protons
\cite{agre:00:nature,schulten:02:science}.

\begin{figure}[tbph]
\includegraphics[width=6.0in]{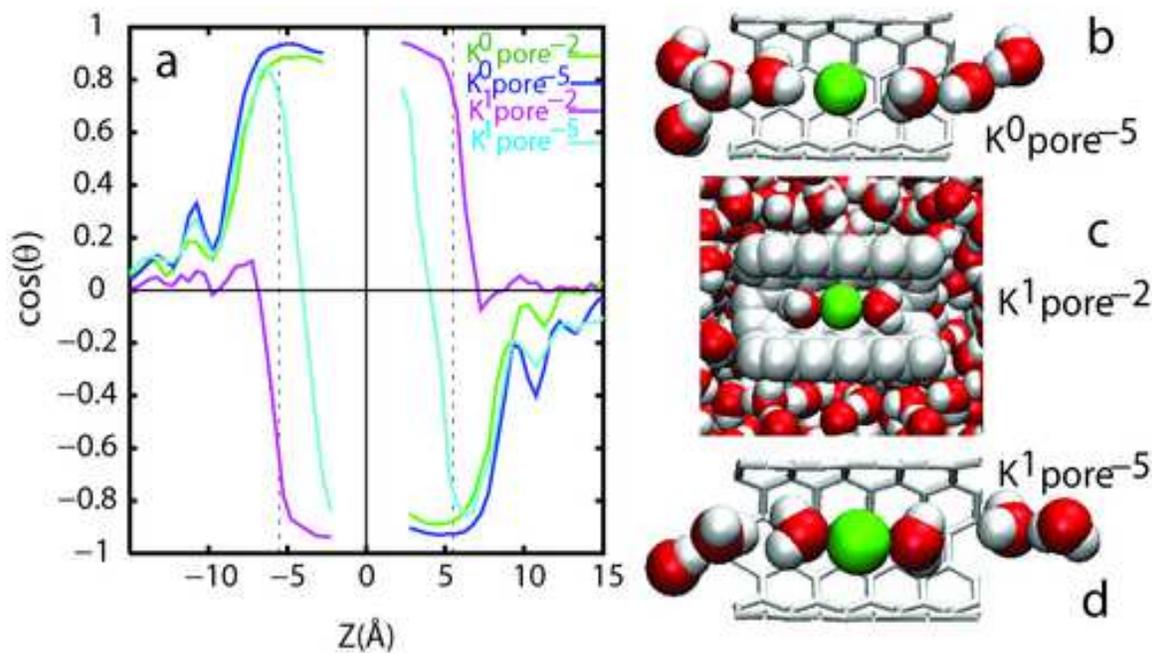}
\caption{Average orientational preferences of water molecules in the
nanopore interior in the presence of a neutral LJ solute ($K^0$) or a
cation ($K^+$) for different charge states of the nanopore. Panels
(b), (c), and (d) show snapshots from simulations.}
\protect\label{fig:orient}
\end{figure}

The low density region at the center of a charged pore is only a few
angstroms wide and presents an ideal location for cations partitioned
into the pore. The lower water density in the cation sized region
ensures lower repulsive interactions without the loss of attractive
interactions with vicinal waters. Indeed, analysis of density
distribution of cations confirms that the most favorable position for
the partitioned cations is at the pore center. The cation-water
electrostatic interactions are strong and affect orientations of the
vicinal waters. Figure \ref{fig:orient} shows the effects of
competitive water-water, water-nanopore, and water-cation interactions
on water orientations.  When the charge on the cation at the pore
center is turned off, {\it e.g.}, for $K^0$Pore$^{-2e}$ case, water
orientations are dictated solely by the water-nanopore electrostatic
interactions.  However, when the charge on the potassium ion is turned
on, {\it i.e.}, for the $K^{+1}$Pore$^{-2e}$ case, the water-potassium
interactions dominate and flip the orientations of two vicinal water
molecules such that their oxygens point toward the $K^+$ ion.
Interestingly, the repulsion between these two water molecules in
ion's shell and others located at the pore ends leads to a low density
area at the pore entry (see Figure \ref{fig:orient}c).  When the
nanopore charge is increased to $-5e$, the water-nanopore interactions
become comparable to water-potassium interactions, and only the two
water molecules vicinal to the $K^+$ ion point their oxygens toward
the ion, and the pore entry region gets filled again (Figure
\ref{fig:orient}d).

{\it Thermodynamics of ion partitioning}: The free energy of ion
transfer from bulk water to nanopore interior quantifies the
thermodynamic driving force for ion partitioning.  The excess free
energy of hydration, $\mu^{ex}_{hyd}$, can be divided into two parts:
$\mu^{ex}_{hyd} = \mu^{ex}_{LJ} + \mu^{ex}_{ele}$, where
$\mu^{ex}_{LJ}$ is the excess chemical potential of hydration of
electrically neutral LJ solutes, and $\mu^{ex}_{ele}$ is the free
energy of charging that solute to its final charge state ($+1e$ here).
We used test particle insertions \cite{widom:63:jcp, widom:82:jpc} of
neutral LJ solutes in bulk water and at the center of a charged
hydrated nanopore to obtain the $\mu^{ex}_{LJ}$ contribution.  As
described previously in detail \cite{Hummer:96:JPC,garde:98:jcp},
calculation of the $\mu^{ex}_{ele}$ was done using the cumulant
expansion method which requires two simulations, one in the uncharged
and another in the fully charged state of the solute in bulk water as
well as in the nanopore interior. Such calculations were performed for
all three cations for nanopore charge of $-2e$, $-5e$, and $-8e$.  For
reference, we also calculated $\mu^{ex}_{LJ}$ and $\mu^{ex}_{ele}$
contributions in a system containing a charged nanopore in vacuum, in
the absence of water molecules.  Comparison of these values to those
in the hydrated nanopore systems provides insights into the role of
water in ion solvation, especially in the confining pore interior
region.

Table \ref{table:hydration} lists values $\mu^{ex}_{LJ}$ and
$\mu^{ex}_{ele}$ for $Na^+$, $K^+$, and $Cs^+$ ions in bulk water, in
the hydrated nanopore interior, and at the center of nanopore in
vacuum for different charge states of the nanopore. In bulk water, the
$\mu^{ex}_{LJ}$ is smallest ({\it i.e.}, most favorable) for $Na^+$
ion, and increases with the ion size. This is expected because the
$\mu^{ex}_{LJ}$ is dominated by repulsive interactions (or the cavity
formation process). Compared to that in bulk water, $\mu^{ex}_{LJ}$
value in the interior of the hydrated nanopore with $-2e$ charge is
smaller for $Na^+$ ion (expected), somewhat higher for $K^+$ ion, and
almost doubles of the $Cs^+$ ion (unexpected).  This is surprising
because we expect that the presence of a low density region in the
hydrated nanopore will reduce, and not increase the $\mu^{ex}_{LJ}$
value for all ions.  The trends become clear, however, when we
consider $\mu^{ex}_{LJ}$ calculated for ions in a nanopore in vacuum.
They indicate that the interior pore of the (5,5) nanopore used here
is small; it is large enough to accommodate $Na^+$ ion comfortably,
but overlaps slightly with the $K^+$ ion, and significantly with the
$Cs^+$ ion.

The advantage arising from the presence of a cavity at the center of
the hydrated nanopores becomes clear when we compare the sum of
$\mu^{ex}_{LJ}$ in bulk water and in bare nanopore to that in hydrated
nanopore. For all ions, that value in hydrated nanopores is smaller
than the sum of bulk water and bare nanopore values. Only for the
highest charge density on the nanopore, the two approach each other as
the central cavity gets gradually filled with water.  Thus, based on
differences between $\mu^{ex}_{LJ}$ in bulk water and in the hydrated
nanopore, we would expect the hydrated nanopore interior to be most
favorable for the smallest $Na^+$ ion and least favorable for the
largest $Cs^+$ ion.  This is contrary to the kinetic behavior observed
in Figure \ref{fig:selectivity}.  To understand the role of
electrostatic interactions and of hydration, below we focus on
differences in $\mu^{ex}_{ele}$ contribution for the three ions.

Table \ref{table:hydration} shows that in bulk water phase,
$\mu^{ex}_{ele}$ is large and negative ({\it i.e.}, favorable) for all
ions.  It is most favorable for $Na^+$ ion, equal to -420 kJ/mol, and
decreases in magnitude as the ion size increases, consistent with
previous calculations \cite{rajamani:04:jcp}. In vacuum, the
electrostatic interaction between cations and the negatively charged
nanopore is highly favorable and is almost identical for all
ions. Minor differences in the energy values in vacuum arise due to
slight differences in the location of ions. Ions are placed at their
most probable locations obtained from ion partitioning data in the
hydrated charged nanopores.  The charging free energy of ions in the
hydrated charged nanopores is also large and negative.  However, its
magnitude is smaller than that in vacuum calculations.  In the charged
and hydrated nanopore, water molecules from either side point their
hydrogen atoms toward the center.  Charging of a neutral LJ solute
located at the center to $+1e$ charge partially reorients those water
molecules to more favorable ion-water configurations.  This effect can
be viewed as ``competitive solvation'', where the cation and the pore
both compete for the same water molecules for their solvation.  As a
result, although the net value of $\mu^{ex}_{ele}$ for ion solvation
is negative, it is smaller in magnitude compared to that inside the
nanopore in vacuum.  Sodium ion, with its small size, and higher
charge density is able to fit in, as well as orient vicinal water
molecules in the pore such that the overall reduction (compared to
that in vacuum) is smaller compared to that for potassium and cesium
ions.

\begin{figure}[tbph]
\includegraphics[width=6.0in]{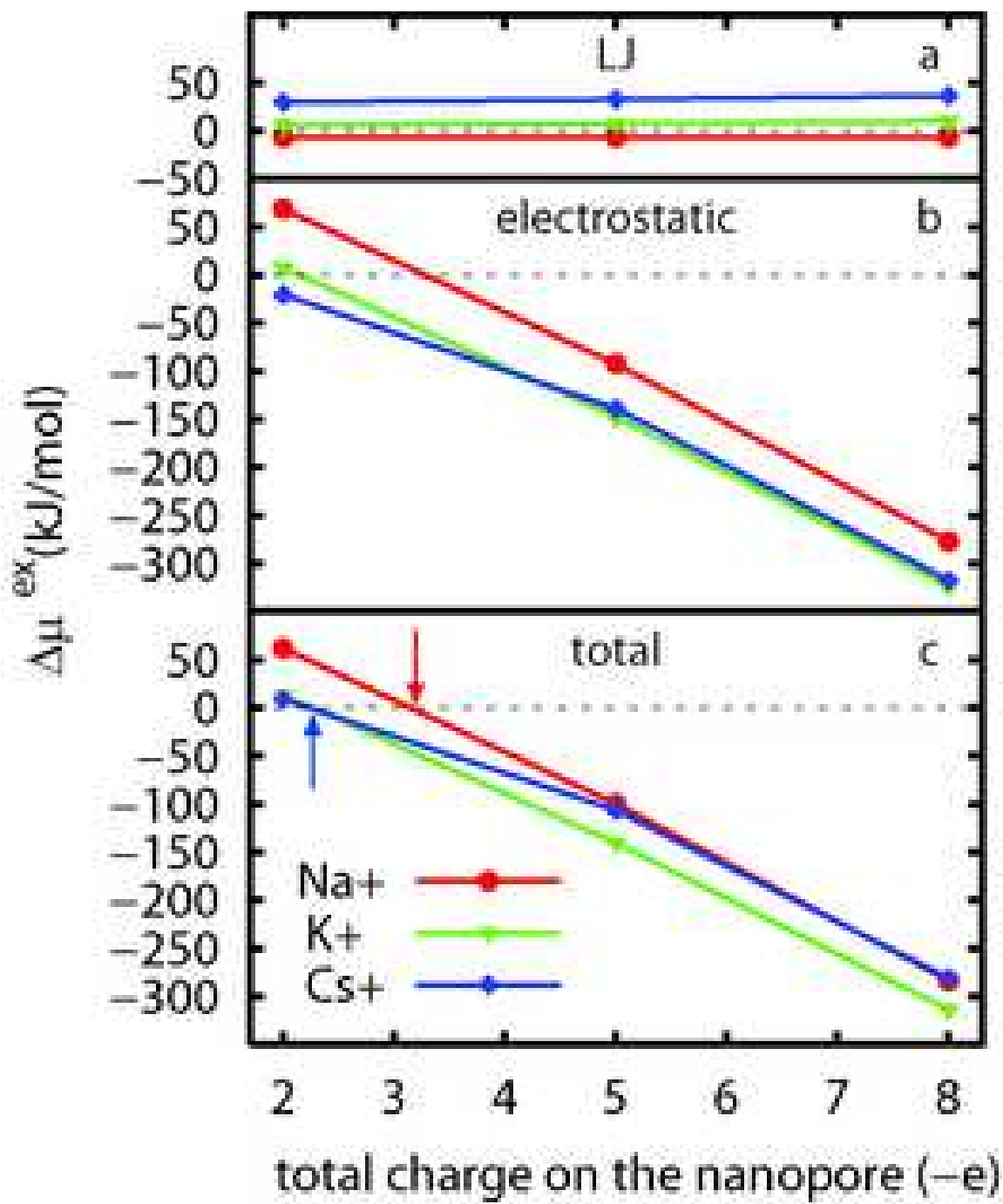}
\caption{Free energy, $\Delta \mu^{ex}_{hyd}$,  of 
water-to-hydrated-nanopore transfer of cations. LJ and electrostatic
contributions to the free energy are also listed. The threshold charge
density at which the transfer free energy becomes favorable is
indicated by arrows.}
\protect\label{fig:dmu}
\end{figure}

The sum, $\mu^{ex}_{hyd}=\mu^{ex}_{LJ}+\mu^{ex}_{ele}$, determines the
overall driving force for ion partitioning into the pore.  The
difference of ion hydration free energy in the hydrated nanopore
interior and in bulk water is the water-to-nanopore transfer free
energy, and is listed in Table \ref{table:transfer} as well as shown
in Figure \ref{fig:dmu} for different charge states of the nanopore.
For $q_{pore}=-2e$, the transfer free energy for all ions is positive
(unfavorable) consistent with the observation from Figure
\ref{fig:selectivity} that none of the ions partitions into the pore
over 12 ns timescale.  For $Na^+$ and $K^+$ ions, the electrostatic
contribution is unfavorable; that is, the overall ion-water-nanopore
interactions do not compensate for the loss of ion-water interactions
in the bulk water.  In contrast, for $Cs^+$ ion, it is the $\Delta
\mu^{ex}_{LJ}$ contribution that is unfavorable as expected from the
somewhat larger size of that ion relative to the nanopore internal
diameter.

Increasing the charge density on the nanopore increases the strength
of cation-nanopore interactions, making the pore interior increasingly
favorable for all three cations.  At a sufficiently high value of
nanopore charge ({\it e.g.}, $q_{pore}=-5e$), the ion-nanopore
electrostatic interactions are large and negative for all ions
indicating that it is thermodynamically favorable for all three types
of ions to partition into the pore.  The free energy varies
approximately linearly with the nanopore charge, and the threshold
value of charge at which the transfer free energy changes from
positive to negative can be estimated from Figure \ref{fig:dmu}. That
threshold charge value is near $-2e$ for $K^+$ and $Cs^+$ ions, and
somewhat larger ($\sim -3e$) for the $Na^+$ ion. That is for nanopore
charge of slightly larger than $-2e$, partitioning of $K^+$ and $Cs^+$
ions will be quite favorable, whereas that of $Na^+$ ion unfavorable,
providing a significant thermodynamic selectivity for partitioning of
those larger ions over the smaller $Na^+$ ion.  At high nanopore
charge densities, although partitioning of all ions into the nanopore
is favorable, the $K^+$ ions appears to benefit from its optimal size
(compared to the nanopore diameter) and corresponding hydration free
energy in water.  The balance of LJ and electrostatic interactions for
$K^+$ ion in bulk water and in the nanopore interior are such that its
partitioning is most favored at higher nanopore charge densities.

Based on the large thermodynamic driving force at charge densities
greater than $q_{pore}=-3e$, we would expect partitioning of all three
ions into the charged nanopores.  Corresponding average partitioning
times are listed in Table \ref{tab:tau}. It is clear that the
timescale for partitioning of $Na^+$ ions for $q_{pore}=-4e$ is
significantly larger than that for $K^+$ and $Cs^+$ ions, which
partition over subnanosecond timescales.  The expected partitioning of
$Na+$ in pores with total charge of $-3e$ would take even longer time,
well beyond the scope of present simulations.  Thus, the thermodynamic
and kinetic data suggest the existence of a free energy barrier for
the ion partitioning process, the height of which is different for
different ions.  Below we discuss possible physical origins of that
barrier.

{\it Possible origins of the free energy barrier based on the
dehydration of ions}: The diameter of the nanopore is small, just
sufficient to accommodate the cations. As seen in Figure
\ref{fig:orient}, only two water molecules are available for 
direct hydration of ions partitioned into the pore. In contrast, the
number of water molecules in the hydration shell of cations in bulk
water is larger, approximately 5.3, 7, and 8, for $Na^+$, K $^+$, and
$Cs^+$ ions, respectively, consistent with previous simulation and
experimental studies \cite{carrillo:03:jcp, rempe:01:fpe,
driesner:98:gca, ohtaki:01:mc}.  Ions thus undergo significant
dehydration as they partition into the nanopore.  As the ion
approaches the end of the nanopore from outside, the ion-nanopore
interactions partially compensate for the loss of ion-water
interactions.  One would require numerous simulations of the
partitioning process or apply a combination of methods such as the
transition path sampling \cite{bolhuis:02:arpc} and umbrella sampling
\cite{torrie:77:jcp,kumar:92:jcp} to obtain detailed structural 
insights into the transition state, the appropriate reaction path, and
the free energy along that path.  We have not performed that analysis
here. However, monitoring fluctuations of the number of hydration
shell water molecules for cations in bulk water provides qualitative
insights into the origin of the free energy barrier.

\begin{figure}[tbph]
\includegraphics[width=6.0in]{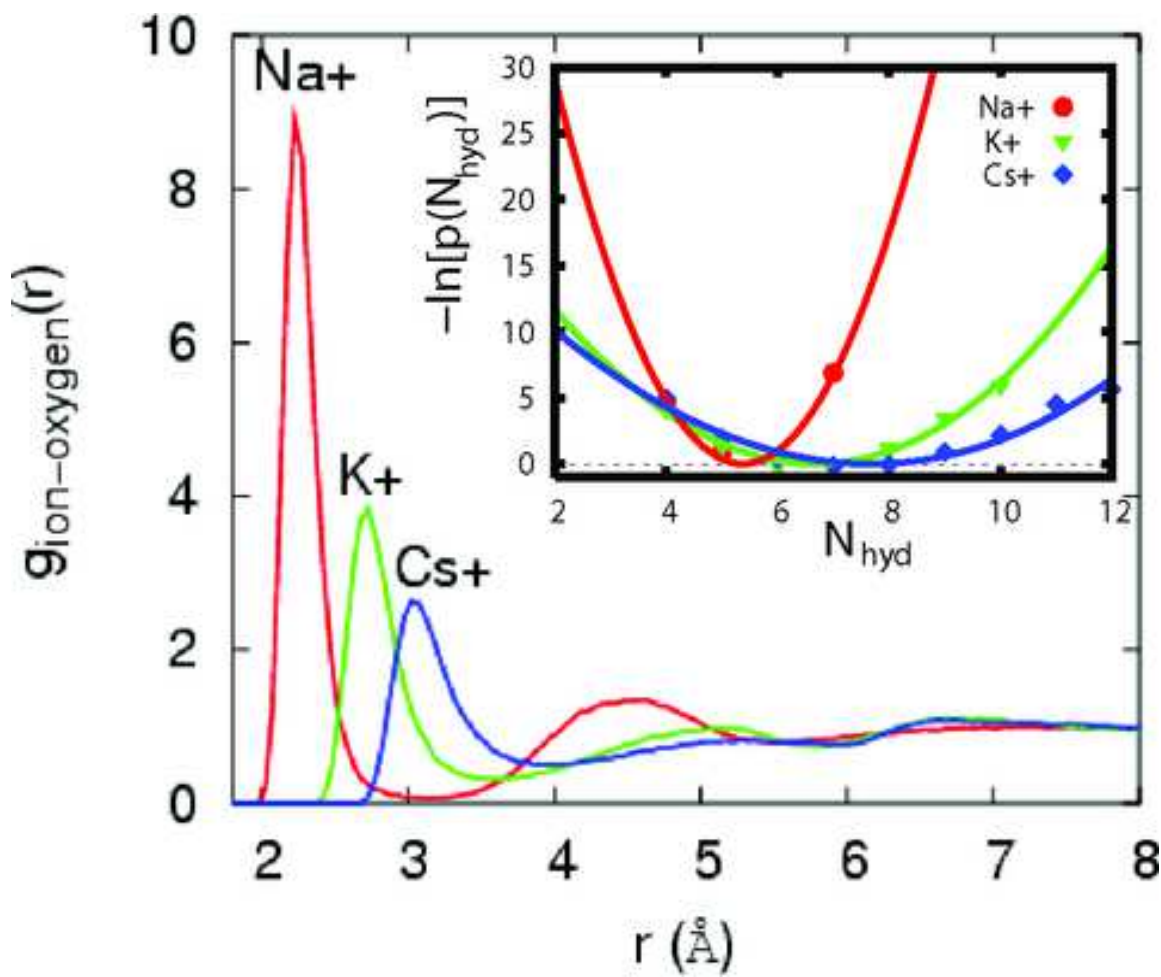}
\caption{Cation-water oxygen radial distribution functions obtained from MD
simulations of cations in bulk water.  The height of the first peak
decreases significantly, indicating weaker binding of water to the ion
with increasing cation size. Inset: Potential of mean force, or the
free energy of spontaneous fluctuations of hydration shell waters in
$kT$ units, $F(N_{hyd})/kT = -{\rm ln}[p(N_{hyd})]$, where
$p(N_{hyd})$ is the probability distribution of number of hydration
shell water molecules.}
\protect\label{fig:dehydration}
\end{figure}

Figure \ref{fig:dehydration} shows the potential of mean force,
-ln$[p(N_{hyd})]$, obtained from probability, $p(N_{hyd})$, of
observing $N_{hyd}$, number of water molecules in the hydration shell
of three different cations.  $N_{hyd}$ fluctuates between 4 and 7 for
$Na^+$ ion, with an average value of 5.3.  The hydration shells of
$K^+$ and $Cs^+$ display higher fluctuations, with $N_{hyd}$ ranging
between 4 to 10 for $K^+$ and 4 to 12 for $Cs^+$, respectively.  The
trends in these fluctuations are consistent with ion-water radial
distribution functions shown in Figure
\ref{fig:dehydration}, which show tighter binding of water molecules to
$Na^+$ compared to that for the larger $K^+$ or $Cs^+$ ions.  The
hydration number of 2 as required in the nanopore interior is never
observed spontaneously in simulations in bulk water, as the free
energy of such dehydration is rather high. However, the estimated
value of free energy for $N_{hyd}=2$ in bulk water, approximately
equal to 10$kT$ for $Cs^+$, 11$kT$ for $K^+$, and over 25$kT$ for
$Na^+$ ion is consistent with the significantly slower partitioning of
$Na^+$ compared to $K^+$ and $Cs^+$ ions.  Despite this consistency,
it is important to note that the dehydration free energies do not
equal the free energy barrier for partitioning of these ions.  The
favorable ion-nanopore interactions assist the dehydration of cations,
therefore reducing the barrier height. This is supported by the fact
that the kinetic selectivity decreases with the increasing of the
nanopore charge density. It will be interesting to investigate the
exact nature of the barrier and especially the role of water structure
at the pore opening in future simulation studies.

\section{Conclusions} 

Nanoporous materials with well defined pore sizes can provide
excellent means to extract or separate small solutes based on steric
or shape effect alone \cite{jirage:97:science, keffer:96:jpc,
heyden:02:ces, ayappa:98:cpl, adhangale:02:langmuir}. However, solvent
can play an important role through differential solvation of solutes
in the bulk and in the pore interior thereby providing selective
partitioning of specific solutes of interest from a mixture of
similarly-sized solutes.  We studied the partitioning of cations
$Na^+$, $K^+$ and $Cs^+$ ions from bulk water into negatively charged
model cylindrical pores with increasing pore charge density using MD
simulations. Neutral and near neutral nanopores exclude ions from
their interior.  However, over a certain range of negative charge
density on the pore, the nanopores display selectivity toward
partitioning of the larger cations $K^+$ and $Cs^+$ over that for the
smaller $Na^+$ ion.  Nonequilibrium kinetic simulations show that the
partitioning is significantly slower for $Na^+$ compared to that for
$K^+$ and $Cs^+$ ions, especially for lower charge densities on the
pore.  Thermodynamic and kinetic observations collectively suggest the
presence of a barrier for partitioning of cations into the nanopore
interior. Such a barrier is expected to slow down the flow rates of
ions through nanopores.  We found that trends in free energy of
dehydration of ions in bulk water are consistent with the partitioning
kinetics.

Although our studies here are focused on a CNT-like model pore, the
results are relevant to studies of carbon nanotubes focused on solute
partitioning and transport \cite{hummer:01:nature, kalra:03:pnas,
waghe:02:jcp, skoulidas:prl:02, sholl:06:science,
aluru:06:nt,Murad:06:jcp, Hansen:jcp:05} and to complementary
theoretical analyzes \cite{piasecki:04:pre, berezhkovskii:02:prl,
chou:04:bj, carrillo:04:prl, daiguji:04:nl, thompson:03:jcp,
ramirez:03:jpc}. Previous studies of neutral hydrated carbon nanotubes
show that water flows through their greasy interior in a frictionless
manner and the flow rate is limited by the entry and exit barriers
\cite{kalra:03:pnas}.  The low density region at the center of charged
nanopores observed here would present an internal barrier for water
transport. It will be interesting to explore to what extent such
internal barriers will affect the flow rates of water through
nanopores.  More broadly, manipulating the balance of various
interactions through design of nanosystems and functionalization
\cite{martin:01:am, mitchell:02:jacs, hinds:05:nature} provides a
mechanism for selective partitioning, separation, or transport of
solutes of interest from a mixture of molecules of similar size and
shape. Lastly, the partitioning of cations into negatively charged
pores of subnanometer dimensions is qualitatively consistent with the
picture emerging from experimental studies of carbon supercapacitors
\cite{chmiola:06:science}.  Simulation studies of more realistic
systems along these lines may aid better design of high energy storage
devices.\\

\vspace{1cm}
\noindent {\bf ACKNOWLEDGMENTS}: SG gratefully acknowledges partial 
financial support of the NSF (CAREER and NSEC) grants, as well as the
NIH RECCR grant. LY thanks hospitality of the Chemical and Biological
Engineering Department at RPI during the Katrina disaster.


\newpage
\begin{table}[t]
\caption{The average, $\left< \tau_1 \right>$, and standard deviation 
$\sigma_{\tau_1}$ of the first passage time (in ns units) for
partitioning of $Na^+$, $K^+$, and $Cs^+$ ions into negatively 
charged nanopores at varying charge densities.}
\protect\label{tab:tau}

\begin{tabular}[t]{ccp{1.5cm}p{1.5cm}p{1.5cm}p{1.5cm}p{1.5cm}}
\hline 
\multicolumn{1}{p{0.5cm}}{} &
\multicolumn{2}{c}{\centering nanopore ($-4e$)} &  
\multicolumn{2}{c}{\centering nanopore ($-5e$)} &
\multicolumn{2}{c}{\centering nanopore ($-6e$)}\\

\multicolumn{1}{c}{} & 
\multicolumn{1}{p{1.5cm}}{\centering $\left< \tau_1 \right>$    } &  
\multicolumn{1}{p{1.5cm}}{\centering $\sigma_{\tau_1}$   } &  
\multicolumn{1}{p{1.5cm}}{\centering $\left< \tau_1 \right>$    } &  
\multicolumn{1}{p{1.5cm}}{\centering $\sigma_{\tau_1}$   } &  
\multicolumn{1}{p{1.5cm}}{\centering $\left< \tau_1 \right>$    } &  
\multicolumn{1}{p{1.5cm}}{\centering $\sigma_{\tau_1}$   } \\
\hline
\multicolumn{1}{c}{\centering $Na^+$    } & 
\multicolumn{1}{c}{\centering 17.46   } &  
\multicolumn{1}{c}{\centering 11.79 } &  
\multicolumn{1}{c}{\centering 3.09 } &
\multicolumn{1}{c}{\centering 2.06  } &  
\multicolumn{1}{c}{\centering 0.92} &  
\multicolumn{1}{c}{\centering 0.65}\\

\multicolumn{1}{c}{\centering $K^+$     } & 
\multicolumn{1}{c}{\centering 0.66   } &  
\multicolumn{1}{c}{\centering 0.50   } &  
\multicolumn{1}{c}{\centering 0.72   } &
\multicolumn{1}{c}{\centering 0.52   } &  
\multicolumn{1}{c}{\centering 0.14   } &  
\multicolumn{1}{c}{\centering 0.04   }\\

\multicolumn{1}{c}{\centering $Cs^+$    } & 
\multicolumn{1}{c}{\centering 0.78      } &  
\multicolumn{1}{c}{\centering 0.41   } &  
\multicolumn{1}{c}{\centering 0.36   } &
\multicolumn{1}{c}{\centering 0.18   } &  
\multicolumn{1}{c}{\centering 0.34   } &  
\multicolumn{1}{c}{\centering 0.34   }\\
\hline 
\end{tabular}
\end{table}

\begin{center}
\begin{table}[t]
\caption{Contributions from LJ ($\mu^{ex}_{LJ}$) and electrostatic 
interactions ($\mu^{ex}_{ele}$) to the free energy of hydration of
cations in bulk water, in hydrated nanopores, and in nanopores in
vacuum (the last three rows) for three different charge states of the
nanopore.  Units: kJ/mol}
\protect\label{table:hydration}
\begin{tabular}[t]{c|cp{1.0cm}p{1.3cm}p{1.3cm}p{1.0cm}p{1.3cm}p{1.3cm}p{1.0cm}p{1.3cm}p{1.3cm}p{1.0cm}p{1.3cm}p{.0cm}}
\hline 
\multicolumn{1}{p{0.6cm}|}{} &  
\multicolumn{3}{c|}{\centering bulk water  } &
\multicolumn{3}{c|}{\centering nanopore (-2e)} &  
\multicolumn{3}{c|}{\centering nanopore (-5e)} &
\multicolumn{3}{c }{\centering nanopore (-8e)} \\
\multicolumn{1}{c|}{} & 
\multicolumn{1}{p{1.0cm}|}{\centering ${\mu}^{ex}_{lj}$    } &  
\multicolumn{1}{p{1.2cm}|}{\centering ${\mu}^{ex}_{ele}$   } &  
\multicolumn{1}{p{1.2cm}|}{\centering ${\mu}^{ex}_{hyd}$ } &
\multicolumn{1}{p{1.0cm}|}{\centering ${\mu}^{ex}_{lj}$    } &  
\multicolumn{1}{p{1.2cm}|}{\centering ${\mu}^{ex}_{ele}$   } &  
\multicolumn{1}{p{1.2cm}|}{\centering ${\mu}^{ex}_{hyd}$ } &
\multicolumn{1}{p{1.0cm}|}{\centering ${\mu}^{ex}_{lj}$    } &  
\multicolumn{1}{p{1.2cm}|}{\centering ${\mu}^{ex}_{ele}$   } &  
\multicolumn{1}{p{1.2cm}|}{\centering ${\mu}^{ex}_{hyd}$ } &
\multicolumn{1}{p{1.0cm}|}{\centering ${\mu}^{ex}_{lj}$    } &  
\multicolumn{1}{p{1.2cm}|}{\centering ${\mu}^{ex}_{ele}$   } &  
\multicolumn{1}{p{1.0cm} }{\centering ${\mu}^{ex}_{hyd}$ } &\\
\hline
\multicolumn{1}{c|}{\centering $Na^+$    } & 
\multicolumn{1}{c|}{\centering 8.9   } &  
\multicolumn{1}{c|}{\centering -420.1} &  
\multicolumn{1}{c|}{\centering -411.2} &
\multicolumn{1}{c|}{\centering 1.9   } &  
\multicolumn{1}{c|}{\centering -351.3} &  
\multicolumn{1}{c|}{\centering -349.4} &
\multicolumn{1}{c|}{\centering 1.8   } &  
\multicolumn{1}{c|}{\centering -512.7} &  
\multicolumn{1}{c|}{\centering -510.9} &
\multicolumn{1}{c|}{\centering 2.0   } &  
\multicolumn{1}{c|}{\centering -696.9} &  
\multicolumn{1}{c }{\centering -694.9} &\\

\multicolumn{1}{c|}{\centering $K^+$     } & 
\multicolumn{1}{c|}{\centering 22.3  } &  
\multicolumn{1}{c|}{\centering -306.9} &  
\multicolumn{1}{c|}{\centering -284.5} &
\multicolumn{1}{c|}{\centering 28.2  } &  
\multicolumn{1}{c|}{\centering -299.3} &  
\multicolumn{1}{c|}{\centering -271.1} &
\multicolumn{1}{c|}{\centering 29.7  } &  
\multicolumn{1}{c|}{\centering -454.6} &  
\multicolumn{1}{c|}{\centering -424.9} &
\multicolumn{1}{c|}{\centering 32.1  } &  
\multicolumn{1}{c|}{\centering -630.3} &  
\multicolumn{1}{c }{\centering -598.3} &\\

\multicolumn{1}{c|}{\centering $Cs^+$    } & 
\multicolumn{1}{c|}{\centering 29.0  } &  
\multicolumn{1}{c|}{\centering -259.7} &  
\multicolumn{1}{c|}{\centering -230.7} &
\multicolumn{1}{c|}{\centering 58.4  } &  
\multicolumn{1}{c|}{\centering -280.1} &  
\multicolumn{1}{c|}{\centering -221.7} &
\multicolumn{1}{c|}{\centering 61.5  } &  
\multicolumn{1}{c|}{\centering -398.6} &  
\multicolumn{1}{c|}{\centering -337.1 } &
\multicolumn{1}{c|}{\centering 65.2  } &  
\multicolumn{1}{c|}{\centering -576.9} &  
\multicolumn{1}{c }{\centering -511.7} &\\
\hline
\multicolumn{4}{l|}{ ~$Na^+$ in nanopore in vacuum $\rightarrow$} & 
\multicolumn{1}{c|}{\centering    -4.8} &  
\multicolumn{1}{c|}{\centering  -442.1} &  
\multicolumn{1}{c|}{\centering  -446.9} &
\multicolumn{1}{c|}{\centering    -4.8} &  
\multicolumn{1}{c|}{\centering -1017.4} &  
\multicolumn{1}{c|}{\centering -1022.2} &
\multicolumn{1}{c|}{\centering    -4.8} &  
\multicolumn{1}{c|}{\centering -1592.1} &  
\multicolumn{1}{c }{\centering -1596.9} &\\

\multicolumn{4}{l|}{ ~$K^+$ ~~in nanopore in vacuum $\rightarrow$} & 
\multicolumn{1}{c|}{\centering    15.1} &  
\multicolumn{1}{c|}{\centering  -441.3} &  
\multicolumn{1}{c|}{\centering  -426.2} &
\multicolumn{1}{c|}{\centering    15.1} &  
\multicolumn{1}{c|}{\centering -1014.3} &  
\multicolumn{1}{c|}{\centering  -999.2} &
\multicolumn{1}{c|}{\centering    15.1} &  
\multicolumn{1}{c|}{\centering -1588.6} &  
\multicolumn{1}{c }{\centering -1573.5} &\\

\multicolumn{4}{l|}{ ~$Cs^+$ ~in nanopore in vacuum $\rightarrow$} & 
\multicolumn{1}{c|}{\centering    43.0} &  
\multicolumn{1}{c|}{\centering  -440.9} &  
\multicolumn{1}{c|}{\centering  -397.9} &
\multicolumn{1}{c|}{\centering    43.0} &  
\multicolumn{1}{c|}{\centering -1014.4} &  
\multicolumn{1}{c|}{\centering  -971.4} &
\multicolumn{1}{c|}{\centering    43.0} &  
\multicolumn{1}{c|}{\centering -1587.3} &  
\multicolumn{1}{c }{\centering -1544.3} &\\

\hline 
\end{tabular}
\end{table}
\end{center}

\begin{table}[t]
\caption{Free energies of water-to-hydrated-nanopore transfer
of cations for three charge states of the nanopore. Contributions from
LJ and electrostatic interactions are also listed. Units: kJ/mol}
\protect\label{table:transfer}
\begin{tabular}[t]{c|cp{1.5cm}p{1.5cm}p{1.5cm}p{1.5cm}p{1.5cm}p{1.5cm}p{1.5cm}p{1.5cm}}
\hline 
\multicolumn{1}{p{0.8cm}|}{} &
\multicolumn{3}{c|}{\centering nanopore (-2e)} &  
\multicolumn{3}{c|}{\centering nanopore (-5e)} &
\multicolumn{3}{c }{\centering nanopore (-8e)} \\
\cline{2-10}
\multicolumn{1}{c|}{} & 
\multicolumn{1}{p{1.5cm}|}{\centering $\Delta{\mu}^{ex}_{lj}$    } &  
\multicolumn{1}{p{1.5cm}|}{\centering $\Delta{\mu}^{ex}_{ele}$   } &  
\multicolumn{1}{p{1.5cm}|}{\centering $\Delta{\mu}^{ex}_{hyd}$ } &
\multicolumn{1}{p{1.5cm}|}{\centering $\Delta{\mu}^{ex}_{lj}$    } &  
\multicolumn{1}{p{1.5cm}|}{\centering $\Delta{\mu}^{ex}_{ele}$   } &  
\multicolumn{1}{p{1.5cm}|}{\centering $\Delta{\mu}^{ex}_{hyd}$ } &
\multicolumn{1}{p{1.5cm}|}{\centering $\Delta{\mu}^{ex}_{lj}$    } &  
\multicolumn{1}{p{1.5cm}|}{\centering $\Delta{\mu}^{ex}_{ele}$   } &  
\multicolumn{1}{p{1.5cm} }{\centering $\Delta{\mu}^{ex}_{hyd}$ }\\
\hline
\multicolumn{1}{c|}{\centering $Na^+$} & 
\multicolumn{1}{c|}{\centering -7.0  } &
\multicolumn{1}{c|}{\centering 68.8  } &  
\multicolumn{1}{c|}{\centering 61.8  } &
\multicolumn{1}{c|}{\centering -7.1  } &  
\multicolumn{1}{c|}{\centering -92.6 } &  
\multicolumn{1}{c|}{\centering -99.7 } &
\multicolumn{1}{c|}{\centering -6.9  } &  
\multicolumn{1}{c|}{\centering -276.8} &  
\multicolumn{1}{c }{\centering -283.7}\\

\multicolumn{1}{c|}{\centering $K^+$     } & 
\multicolumn{1}{c|}{\centering 5.9   } &  
\multicolumn{1}{c|}{\centering 7.6   } &  
\multicolumn{1}{c|}{\centering 13.5  } &
\multicolumn{1}{c|}{\centering 7.4   } &  
\multicolumn{1}{c|}{\centering -147.7} &  
\multicolumn{1}{c|}{\centering -140.3} &
\multicolumn{1}{c|}{\centering 9.8   } &  
\multicolumn{1}{c|}{\centering -323.4} &  
\multicolumn{1}{c }{\centering -313.6}\\

\multicolumn{1}{c|}{\centering $Cs^+$    } & 
\multicolumn{1}{c|}{\centering 29.4  } &  
\multicolumn{1}{c|}{\centering -20.4 } &  
\multicolumn{1}{c|}{\centering 9.0   } &
\multicolumn{1}{c|}{\centering 32.5   } &  
\multicolumn{1}{c|}{\centering -138.9} &  
\multicolumn{1}{c|}{\centering -106.4} &
\multicolumn{1}{c|}{\centering 36.2  } &  
\multicolumn{1}{c|}{\centering -317.2} &  
\multicolumn{1}{c }{\centering -281.0}\\
\hline 
\end{tabular}
\end{table}
\clearpage

\newpage

\newpage

\newpage

\newpage

\newpage

\end{document}